# Revealing the Biexciton and Trion-exciton Complexes in BN Encapsulated WSe$_2$

Zhipeng Li[1,2,#], Tianmeng Wang[1,#], Zhengguang Lu[3,4], Chenhao Jin[5], Yanwen Chen[1], Yuze Meng[1,6], Zhen Lian[1], Takashi Taniguchi[7], Kenji Watanabe[7], Shengbai Zhang[8], Dmitry Smirnov[3], Su-Fei Shi[1,9*]

1. Department of Chemical and Biological Engineering, Rensselaer Polytechnic Institute, Troy, NY 12180
2. School of Chemistry and Chemical Engineering, Shanghai Jiao Tong University, Shanghai, 200240, China
3. National High Magnetic Field Lab, Tallahassee, FL, 32310
4. Department of Physics, Florida State University, Tallahassee, Florida 32306, USA
5. Physics Department, University of California, Berkeley, CA 94720
6. College of Physics, Nanjing University, Nanjing, 210093, P. R. China
7. National Institute for Materials Science, 1-1 Namiki, Tsukuba 305-0044, Japan.
8. Department of Physics, Applied Physics, and Astronomy, Rensselaer Polytechnic Institute, Troy, NY 12180
9. Department of Electrical, Computer & Systems Engineering, Rensselaer Polytechnic Institute, Troy, NY 12180
[#] These authors contributed equally to this work
[*] Corresponding author: shis2@rpi.edu

**Strong Coulomb interactions in single-layer transition metal dichalcogenides (TMDs) result in the emergence of strongly bound excitons, trions and biexcitons. These excitonic complexes possess the valley degree of freedom, which can be exploited for quantum optoelectronics. However, in contrast to the good understanding of the exciton and trion properties, the binding energy of the biexciton remains elusive, with theoretical calculations and experimental studies reporting discrepant results. In this work, we resolve the conflict by employing low-temperature photoluminescence spectroscopy to identify the biexciton state in BN encapsulated single-layer WSe$_2$. The biexciton state only exists in charge neutral WSe$_2$, which is realized through the control of efficient electrostatic gating. In the lightly electron-doped WSe$_2$, one free electron binds to a biexciton and forms the trion-exciton complex. Improved understanding of the biexciton and trion-exciton complexes paves the way for exploiting the many-body physics in TMDs for novel optoelectronics applications.**

Single layer transition metal dichalcogenides (TMDs) represent a new class of atomically thin semiconductors with superior optical and optoelectronic properties[1,2]. The two-dimensional nature of single-layer TMDs results in reduced screening and enhanced Coulomb interaction, giving rise to excitonic complexes such as exciton, trion, and biexciton with binding energy orders of magnitude larger than that of conventional semiconductors such as GaAs[3]. Large spin-orbit coupling[4–7] leads to the splitting of the valence bands in TMDs, and the resulting valence band minimum with different spin configurations breaks the symmetry at the corners of TMDs' Brillouin zone, i.e., K and K' valleys. The different valleys can be accessed selectively through circularly polarized light[8–10], providing a valley degree of freedom that can be exploited for valleytronics[11–14]. Besides the exciton, the higher order charge-complexes such as trions[15,16] and biexcitons also possess a valley degree of freedom, and they can be further used as entangled photon sources[17]. Fundamental understanding of the charge-complexes in TMDs is thus critical for fully utilizing the potential of TMDs for novel optoelectronics applications.

Despite recent developments in the understanding of excitons and trions in TMDs[18,19], an unambiguous measure of the biexciton binding energy remains elusive. Many experimental studies based on linear photoluminescence (PL) spectroscopy reported a biexciton binding energy larger than theoretical predictions[20–27], while recent coherent pump-probe spectroscopy measurements reported a value close to the theoretical prediction[28,29]. Here we demonstrate that, by fabricating a clean single layer $WSe_2$ sandwiched by two h-BN flakes, we can detect an unambiguous evidence of biexciton PL emission in charge neutral $WSe_2$ at low temperature, and the observed binding energy is in excellent agreement with the theoretical calculations[30–34]. We also perform PL spectroscopy as a function of the carrier density and reveal the formation of the electron-bound biexciton exists in the lightly n-doping regime only.

To preserve the $WSe_2$ monolayer crystal quality and probe the intrinsic optical behavior, we constructed the BN/$WSe_2$/BN structure with the pickup method[35], during which the $WSe_2$ was never exposed to any polymer (Section S1 of SI). A piece of few-layer graphene (labeled as graphite) was used as the contact electrode of the single layer $WSe_2$, while another piece was used as the transparent top-gate electrode on the top layer BN, as shown in Fig. 1a and 1b. The high quality of the $WSe_2$ sample is illustrated in Fig. 1c, which displays low-temperature PL spectra measured under the 633 nm continuous-wave (CW) laser excitation. Indeed, the linewidth of the exciton (located at 1.740 eV) can be as narrow as ~ 4 meV, significantly less than that from the typical single-layer exfoliated $WSe_2$ on $SiO_2$[36,37], which results in a clear fine structure of well separated PL features. The emission from the exciton ($X_0$) is observed at 1.740 eV. The two peaks at 1.705 eV and 1.712 eV correspond to intravalley and intervalley trions[38,39], which we label as $X_1^-$ and $X_2^-$, respectively. The PL peak at 1.697 eV is attributed to the emission from the spin-forbidden dark exciton transition[40–42], which is observed here because our high numerical aperture (NA) objective collects the out-of-plane *p*-polarized dark exciton emission[43].

The most salient feature of the PL spectra in Fig. 1c is that, as the excitation power increases, additional PL peaks at 1.723 eV and 1.691 eV emerge. The integrated PL intensity of these two peaks, along with the exciton PL intensity, is plotted as a function of the excitation power in Fig. 1d. The power law dependence of the integrated PL intensity, expressed as $I \propto P^\alpha$, where $I$ stands for the integrated PL intensity and $P$ for the excitation power, is evidently different for the two new emergent peaks compared with that of the exciton. From the fitting to the experimental data in Fig.

1c, we obtained α=1.94 for the peak at 1.723 eV and α=1.82 for the peak at 1.691 eV, significantly larger than that of the exciton (α=1.29) and close to what is expected for biexciton (α=2).

The binding energy of the biexciton, $\Delta_{XX}$, is defined as the energy difference between the two excitons in the free state and the biexciton in the bound state. If we assume that the radiative recombination of the biexciton emits one photon, $\Delta_{XX}$ is equal to the emitted photon energy of the free exciton ($\hbar\omega_X$) minus the emitted photon energy of the biexciton ($\hbar\omega_{XX}$), i.e. $\Delta_{XX} = \hbar\omega_X - \hbar\omega_{XX}$. Theoretical calculations have predicted the biexciton binding energy to be ~20 meV[33,34,44,45]. The emerging PL peak at 1.723 eV lies 17 meV below the exciton peak, in excellent agreement with the predictions for the biexciton binding energy. Considering the nearly quadratic (α=1.94) power dependence of the PL intensity, we assign the peak at 1.723 eV to the true biexciton peak which we label as XX. The other emerging PL peak at 1.691 eV, 49 meV below the exciton peak, was recently reported as the biexciton peak[20]. Considering that the power law exponent α=1.82, being close to the expected value of 2 for a biexciton, we interpret this peak as a negative charge bound biexciton $XX^-$, based on the following PL spectra study as a function of the gate voltage. We note that the XX and $XX^-$ PL peaks are detected at as low excitation power as 40 μW, which corresponds to a power density of 1274 W/cm$^2$ and an exciton density of $6 \times 10^9$ cm$^{-2}$ for a CW laser excitation centered at 633 nm (assuming that the absorption coefficient of WSe$_2$ is 10% and the lifetime of bright exciton is around 15 ps[2,46,47]), significantly lower than what has been used to observe the previous "biexciton" peak, which either involves pulse laser excitation[20] or CW laser excitation with large power[25].

To investigate further the origin of the emerging PL peaks, we measured the PL spectra as a function of the carrier density using another device with an efficient top gate enabling accurate tuning between the p-type and n-type doping regimes. The gate-voltage dependent PL spectra are shown as a color plot in Fig. 2a, in which the color represents the PL intensity. The PL spectra at specific gate voltages, corresponding to the line cuts in Fig. 2a, are shown in Fig. 2b. At the top gate voltage of -0.5 V (black line in Fig. 2b), the WSe$_2$ is close to charge-neutral, and only two PL peaks are observed. The exciton (X) peak is the most pronounced one centered at 1.724 eV, and the biexciton (XX) peak is located at 1.708 eV, thus indicating the 16 meV biexciton binding energy for the second device. At the gate voltage of -2.5 V (magenta line in Fig. 2b), the WSe$_2$ is strongly p-doped and the positive trion peak $X^+$ occurs at 1.701 eV, while the exciton peak disappears. At the top gate voltage of 0.2 V (blue line in Fig. 2b), the WSe$_2$ is n-doped, and the exciton peak is quenched. However, the two negative trion peaks emerge at 1.689 eV and 1.696 eV, which corresponds to the intravalley and intervalley trions, respectively[39]. Interestingly, the biexciton (XX) peak disappears, but the $XX^-$ peak at 1.676 eV appears.

The sensitive gate dependence of the XX and $XX^-$ peaks closely correlates to the PL intensity of the exciton and trions. We plot the integrated PL intensity as a function of the gate voltage for different peaks in Fig. 2c. It is evident that the PL intensity of the exciton peak (black line in Fig. 2c) quickly vanishes in the strongly p-doped ($V_{TG}$ < -2.0 V) and n-doped ($V_{TG}$ > 0.5 V) regimes. The PL for the trions, both the positive and the negative ones, is sensitive to the gate voltage. The PL for the positive trion ($X^+$, blue line) occurs when the gate voltage is less than -1.0 V, and the PL for the negative trions emerges when the gate voltage is greater than -0.5 V. This observation enables us to determine the charge-neutral region to be between -1.0 to -0.5 V, where the biexciton PL is peaked (red line in Fig. 2c). In contrast, the PL of XX⁻ (purple line in Fig. 2c) does not show up in the charge-neutral region, and only emerges at higher gate voltage from -0.5 V to 0.5 V (corresponding to the slightly n-doped region). Considering the doping dependence of XX⁻ PL and its characteristic intensity power law, we assign XX⁻ to an electron-bound biexciton.

The detailed valley configuration of XX and XX⁻ can be probed using the circularly polarized PL spectroscopy, where we selectively excite and probe a particular valley of WSe$_2$. The band structure at the K and K' valleys for WSe$_2$ is shown in Fig. 3b. It is worth noting that the spin-orbit coupling not only gives rise to the large spin splitting in the valence band but also generates a splitting on the order of tens of meV in the conduction band[4–7,48,49]. For WS$_2$ and WSe$_2$, this splitting leads to an opposite spin configuration between the conduction band minimum (CBM) and the valence band maximum (VBM) in the same valley, hosting a ground state exciton which is optically dark[40,42]. We excited the device with circularly polarized light, such as right circularly polarized light ($\sigma^+$), and detected the PL emission with the same or opposite helicity. The valley polarization, defined as the ratio of right circular and left circular components of the emitted PL, i.e., $P = \frac{I_{\sigma^+} - I_{\sigma^-}}{I_{\sigma^+} + I_{\sigma^-}}$, measures the capability of the TMD to maintain the valley information. Interestingly, XX and XX⁻ exhibit higher valley polarization of 0.19 and 0.20, respectively, both higher than the valley polarization of the exciton (0.12), as shown in Fig. 3a. This valley polarization suggests that both XX and XX⁻ involve a bright exciton in the K valley, which we excited with $\sigma^+$ light. Since XX⁻ is a five-particle complex, the lowest energy configuration that maintains the valley polarization can only be the one shown in Fig. 3b. This configuration can be viewed as a biexciton bound to a free electron, or equivalently, a trion bound to one dark exciton. The negative charge bound biexciton is thus effectively a trion-exciton complex (negative trion). The biexciton state of XX is composed of four particles, and one possible configuration is shown in Fig. 3c, with one bright exciton in K valley and one dark exciton in K' valley. One alternative configuration is shown in the SI (Fig. S2e). Our following magneto-PL spectroscopy study confirms that the configuration shown in Fig. 3c is the right configuration for XX.

In the presence of an out-of-plane magnetic field, the degeneracy of the K and K' valleys is lifted, and the spectra of K (K') valley can be selectively accessed through $\sigma^+$ ($\sigma^-$) excitation with $\sigma^+$ ($\sigma^-$) detection configuration as shown in Fig. 4d. The PL spectra as a function of the magnetic field for different valleys are shown in Fig. 4a ($\sigma^-\sigma^-$) and in Fig. 4b ($\sigma^+\sigma^+$) as a color plot. It is evident that the PL peak positions (dashed lines) in Fig. 4a and 4b exhibit a linear shift as a function of the B field due to the Zeeman shift, and the slope is opposite in sign for the $\sigma^+\sigma^+$ and $\sigma^-\sigma^-$ configurations. The opposite Zeeman shifts for the different valleys result in a Zeeman splitting of the PL peaks in the spectra, as shown in Fig. 4d. The Zeeman splitting between the two valleys can be expressed as $\Delta = \Delta_{\sigma^+\sigma^+} - \Delta_{\sigma^-\sigma^-} = g\mu_B B$, where the $\mu_B$ is the Bohr magneton, which is about 58 μeV/T, and $\Delta_{\sigma^+\sigma^+}$ ($\Delta_{\sigma^-\sigma^-}$) is the Zeeman shift in the $\sigma^+\sigma^+$ ($\sigma^-\sigma^-$) configuration. Linear fitting of the experimentally obtained Zeeman splitting $\Delta$ as a function of the B field (Fig. 4c) determines that the g-factor for the exciton (black dots in Fig. 4c) is $-3.64 \pm 0.08$. Our experimentally extracted value is in good agreement with the theoretical calculation $g_X = -4.0$ (Table S1 in SI), and it is also consistent with previous reports[7,50,51]. The PL emission of XX and XX⁻ involves the recombination of one bright exciton, and the expected g factor should be the same as that for the exciton. We obtained a XX $g$-factor (Fig. 4c) of $-4.03 \pm 0.07$ and a XX⁻ $g$-factor of $-5.33 \pm 0.18$ in Fig. 4c, also in good agreement with the theoretical expectation. The deviation from the theoretical expectations is potentially due to the increased Coulomb interaction in the many-particle complex, which is not taken into account in the theoretical calculations[30,31].

The PL intensity ratio of the two Zeeman splitted states ($\sigma^+\sigma^+$ and $\sigma^-\sigma^-$) in the B field provides insight into the detailed configuration of the biexciton. The Zeeman splitting lifts the energy degeneracy at the K and K' valleys and creates two states with the energy difference of $\Delta = g\mu_B B$, which leads to unequal complex population of the two states, and hence, different PL emission intensities. We use the quasi thermal equilibrium picture for a qualitative understanding of the PL difference for the $\sigma^+\sigma^+$ and $\sigma^-\sigma^-$ configurations. The exciton PL intensity in the B field, shown in Fig. 4d, is higher in the state that emits photon with the lower energy ($\sigma^-\sigma^-$), consistent with the quasi-thermal equilibrium picture. We emphasize, however, a quantitative understanding is not easily accessible because of the short exciton lifetime and the finite intervalley scattering complicates the exact thermal equilibrium of the two states. In contrast to the behavior of the exciton, the spectrally higher energy state for biexciton (XX) at ~ 1.726 eV exhibits much higher PL intensity than that of the lower energy state in the spectra, as shown in Fig. 4d. The apparent contradiction arises from the fact that the real energy difference between the two Zeeman splitted states for the biexciton (or other high order excitonic complex) is determined by the total $g$-factor, $g^t$, which should include the contributions from all the constituent particles. On the contrary, the emission photon energy in PL is only determined by the bright exciton which radiatively recombines to emit a photon, and hence the PL peak position difference between the two states is determined by the spectral $g$-factor, $g^s$. The difference between the total g factor ($g^t$) and the spectral $g$-factor ($g^s$) indicates that the spectrally lower energy state is not necessarily lower in total energy.

We can calculate the total $g$-factor for the biexciton to be $g^t_{XX} = 4.0$ for the configuration in Fig. 3c (see SI), which has the opposite sign of the spectral $g$-factor of $g^s = -4.03 \pm 0.07$ (experiment) or $-4.0$ (theory). The positive total $g$-factor means that the spectrally higher energy state (PL peak emits higher energy photon) actually has overall lower energy and should have higher PL intensity. This idea is consistent with our observation in Fig. 4d (XX panel). The alternative configuration involving the intervalley dark exciton has a total $g$-factor of zero (see SI), and it cannot explain the drastic biexciton PL intensity difference between the $\sigma^+\sigma^+$ and $\sigma^-\sigma^-$ configurations. The total $g$-factor for the XX⁻ state is calculated to be 6.0 (see SI), which is also opposite in sign to the spectral $g$-factor of $-5.33 \pm 0.18$ (experiment). Therefore, for the XX⁻ state, $\sigma^-\sigma^-$ configuration is also lower in total energy and should have higher PL intensity, which is consistent with our observation (Fig. 4d).

It is interesting to note that the magneto-PL spectra of the dark exciton state at 1.691 eV are intrinsically different from other PL peaks. While all other states exhibit a single peak in the circularly polarized detection scheme, the emission from the dark exciton splits into two peaks in the B field (magenta lines in Fig. 4a, b). This is because that the optical selection rule in TMDs originates from the conservation of the out-of-plane quasi angular momentum with respect to the three-fold rotational symmetry. However, the optically dark exciton can only couple to an out-of-plane optical field, which has zero out-of-plane angular momentum. Therefore, responses from both valleys can be excited and detected simultaneously, which was detected by an objective with large NA and show up as the two separate peaks in the spectra. The spectra *g*-factor of the dark exciton is experimentally determined to be $-9.75 \pm 0.18$ (Fig. 4c), in good agreement with the expected value of -8.0 (see SI) and the experimental result with the g factor of $-9.4 \pm 0.1$[52]. The unique behavior of the dark exciton confirms that the emission from the excitonic complexes of biexciton and trion-exciton both originate from a bright exciton.

In summary, we have fabricated a high-quality single-layer $WSe_2$ device by employing BN encapsulation and revealed both the biexciton and (negative) trion-exciton complexes through low-temperature PL spectroscopy. The binding energy of the biexciton in the BN encapsulated single layer $WSe_2$ is determined to be about 16-17 meV, in agreement with the theoretical calculations[30,31] and previous coherent pump-probe measurements[28,29]. The biexciton state only exists in charge neutral $WSe_2$, and the trion-exciton complex only emerges in lightly n-doped $WSe_2$. Improved understanding of the high-order excitonic complexes in high-quality $WSe_2$ devices will enable further investigation of many-body physics and furnish new opportunities for novel quantum optoelectronics based on TMDs.

**Acknowledgement**

The device fabrication was supported by Micro and Nanofabrication Clean Room (MNCR), operated by the Center for Materials, Devices, and Integrated Systems (cMDIS) at Rensselaer Polytechnic Institute (RPI). Su-Fei Shi acknowledges the support from Rensselaer Polytechnic Institute and the Center for Future Energy Systems (CFES), a New York State Center for Advanced Technology at RPI. We thank Prof. Feng Wang, Prof. Ronald Hedden, Prof. Jun Yan, Dr. Long Ju, Ting Cao, and Debjit Ghoshal for helpful discussions.

**Contributions**

S.-F. Shi conceived the experiment. Z. Li, Y. Meng, Y. Chen and Z. Lian fabricated the devices. Z. Li, T. Wang and Z. Lu performed the measurements. S.-F. Shi, Z. Li, T. Wang, C. Jin and Y. Chen analyzed the data. S.-F. Shi supervised the project. S.-F. Shi wrote the manuscript with the input from all the other co-authors. All authors discussed the results and contributed to the manuscript.

**Competing financial interests**: The authors declare no competing financial interests.

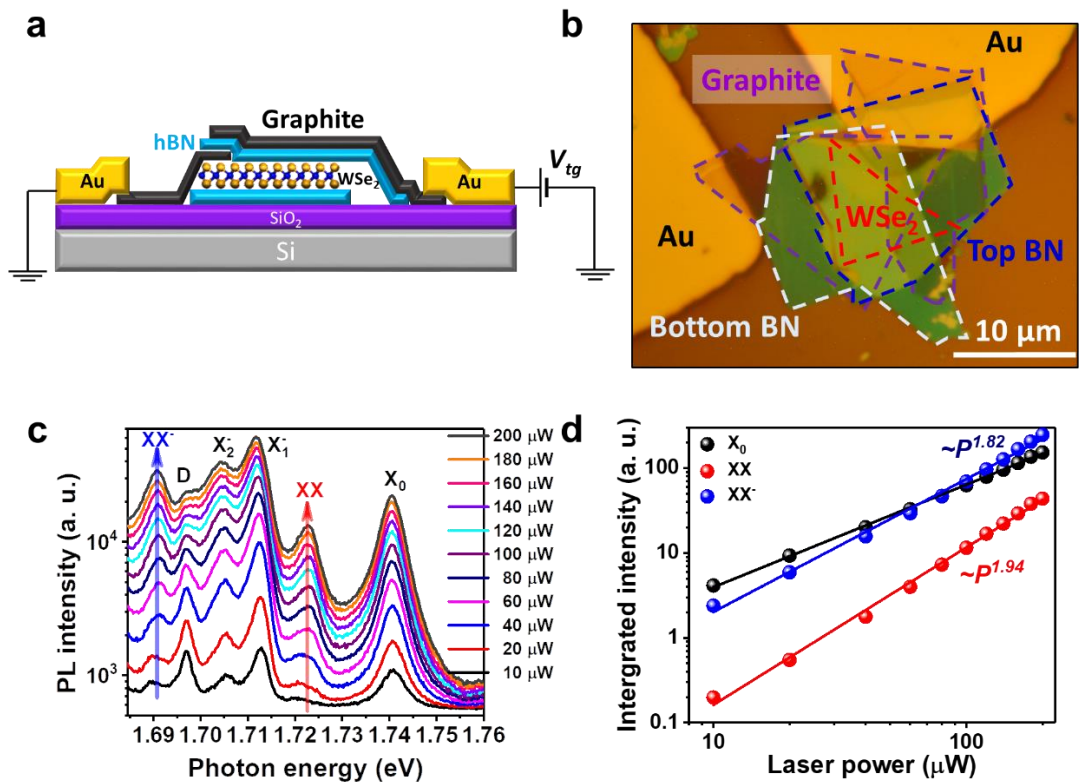

**Figure 1. PL spectra of BN encapsulated single layer WSe₂ device at 4.2 K.** (a) Schematic of the BN encapsulated single layer WSe₂. One piece of few-layer graphene (graphite) is used as the contact electrode and another piece is used as the transparent top gate electrode. (b) The microscope image of the device. Scale bar: 10 μm. (c) PL spectra of the single layer WSe₂ for different excitation powers. The CW laser of wavelength 633 nm was used as the excitation source. (d) Integrated PL intensity of WSe₂ as a function of the excitation power, and the XX and XX⁻ peaks clearly exhibit a nonlinear power dependence with the power law close to 2.

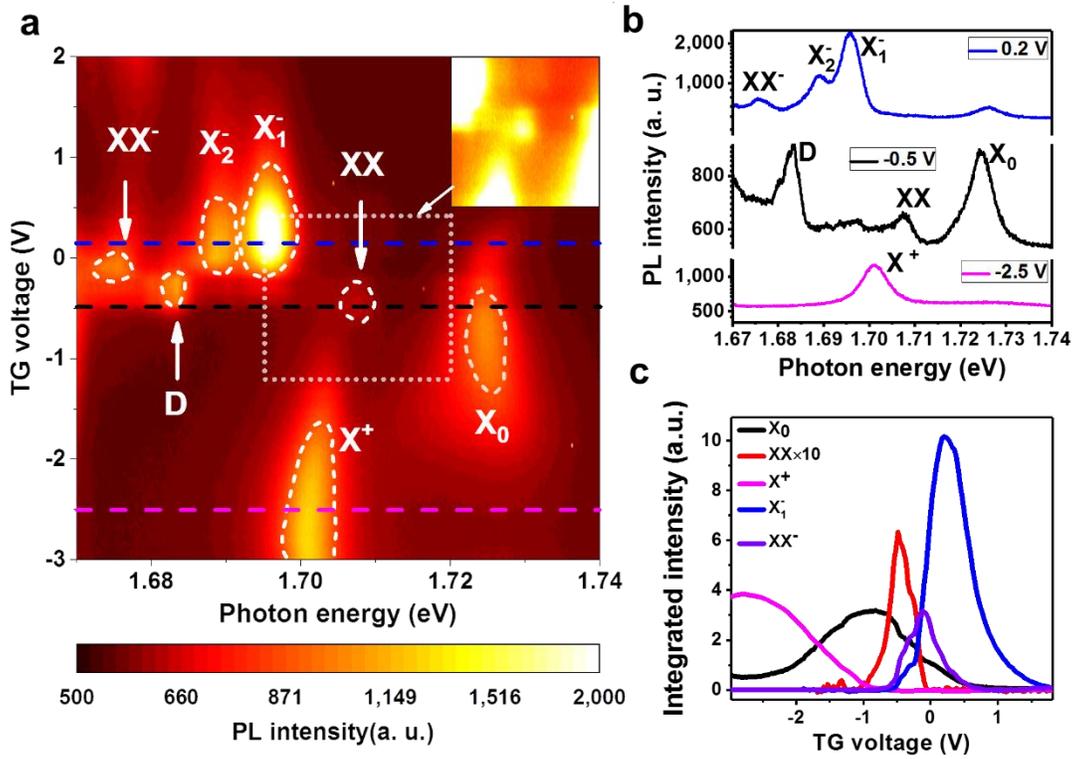

**Figure 2. PL spectra of WSe$_2$ as a function of the top gate voltage.** (a) Color plot of the PL spectra as a function of the top gate voltage. The color represents the PL intensity. Inset: enhanced color plot of the biexciton region. (b) PL spectra at the top gate voltage of -2.5 V, -0.5 V and 0.2 V, corresponding to p-doped (magenta), intrinsic (black), and n-doped (blue) region. (c) Integrated PL intensity for $X_0$, XX, $X^+$ and $X_1^-$ as a function of the top gate voltage. It is evident that the XX only exists in the charge neutral region while XX¯ exits in the lightly n-doped region.
11

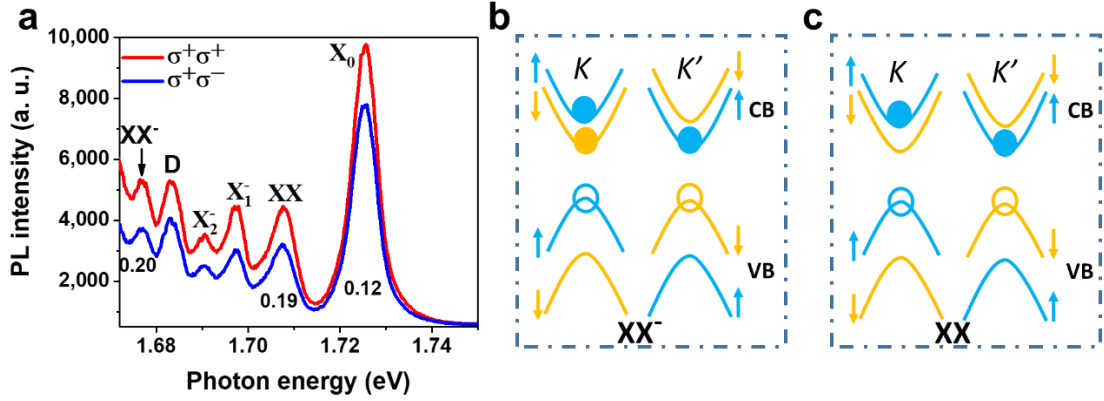

**Figure 3. Valley polarized PL spectra and biexciton configurations.** (a) PL spectra with circularly polarized ($\sigma^+$) excitation and the same ($\sigma^+$) or opposite ($\sigma^-$) helicity detection. Valley polarization, defined as $\frac{I_{\sigma^+} - I_{\sigma^-}}{I_{\sigma^+} + I_{\sigma^-}}$, where $I$ is the integrated PL intensity, is calculated for exciton (0.12), biexciton (0.19) and electron bound biexciton (0.20). (b-c) trion-exciton and biexciton configurations with the solid dots being the electron and the empty one being the hole. Blue color stands for spin up and orange for spin down.

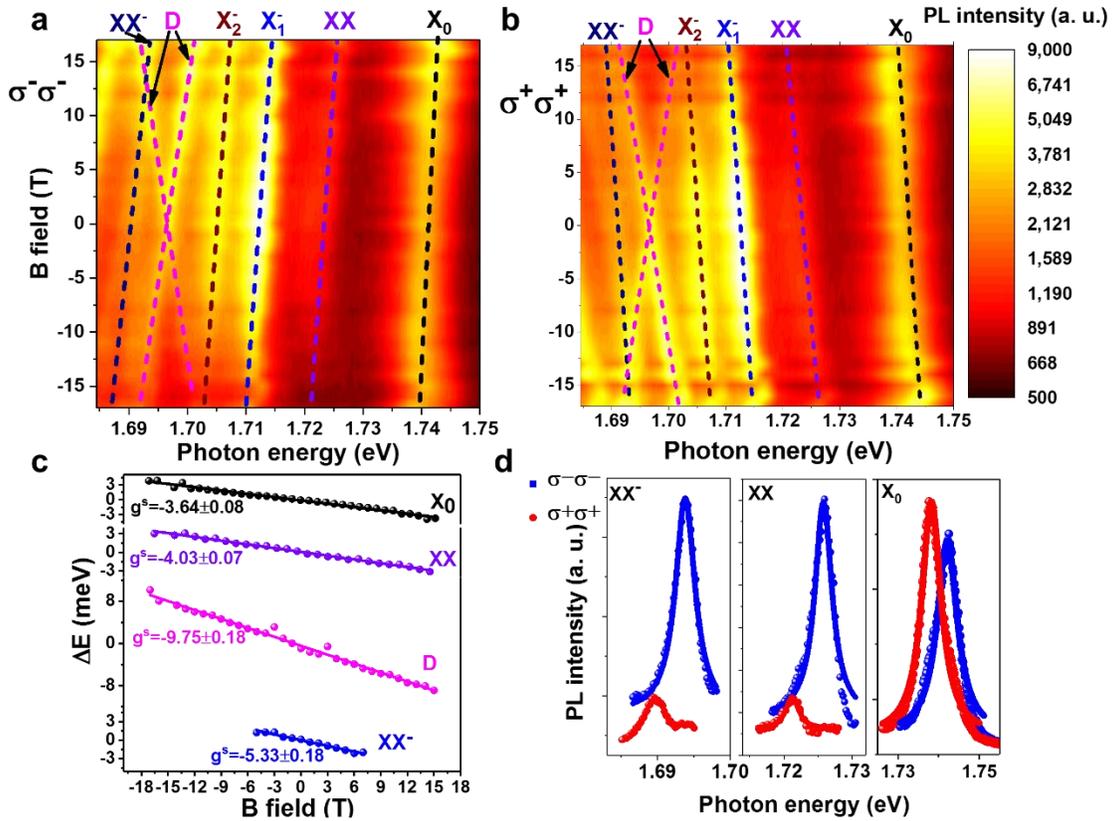

**Figure 4 Magneto-PL spectra of WSe$_2$.** (a) and (b) Color plot of the PL spectra of WSe$_2$ as a function of B field at 4.2 K for the $\sigma^-\sigma^-$ and $\sigma^+\sigma^+$ configuration. The dashed lines are the eye guide to the shift of different PL peaks (c) *g*-factors for different peaks calculated from the Zeeman splitting between the $\sigma^-\sigma^-$ and $\sigma^+\sigma^+$ states as a function of B field, extracted from the magneto-PL spectra in (a) and (b). (d) PL spectra for $\sigma^+(\sigma^-)$ excitation and $\sigma^+(\sigma^-)$ detection of exciton, biexciton and trion-exciton complexes, respectively (B =17 T).

13